\documentclass[aps,prl,reprint,floatfix]{revtex4-2}

\usepackage{amssymb}
\usepackage{amsmath}
\usepackage{amsfonts}
\usepackage{graphicx}

\newcommand{\FE}{\kappa}

\renewcommand{\vec}[1]{\boldsymbol{#1}}

\newcommand{\beq}{\begin{eqnarray}}
\newcommand{\eeq}{\end{eqnarray}}

\newcommand{\tr}{\text{Tr}}
\newcommand{\Tr}{\text{Tr}}

\newcommand{\half}{\frac{1}{2}}

\newcommand{\Hx}{\text{Hx}}

\newcommand{\xrm}{\text{x}}
\newcommand{\crm}{\text{c}}
\newcommand{\Hrm}{\text{H}}
\newcommand{\xc}{\text{xc}}
\newcommand{\Hxc}{\text{Hxc}}

\newcommand{\rxc}{\text{rxc}}

\newcommand{\KS}{\text{KS}}
\newcommand{\GKS}{\text{GKS}}

\newcommand{\MGKS}{\text{MGKS}}

\newcommand{\pr}{^{\prime}}

\renewcommand{\vr}{\vec{r}}

\newcommand{\vrp}{\vec{r}\pr}

\newcommand{\ket}[1]{\left|#1\right\rangle}

\newcommand{\iket}[1]{|#1\rangle}
\newcommand{\ibraket}[2]{\langle#1|#2\rangle}
\newcommand{\ibraketop}[3]{\langle#1|#2|#3\rangle}

\newcommand{\HF}{{\text{HF}}}

\newcommand{\DFA}{{\text{DFA}}}

\newcommand{\up}{\mathord{\uparrow}}
\newcommand{\down}{\mathord{\downarrow}}

\newcommand{\vh}{\hat{v}}
\newcommand{\nh}{\hat{n}}
\newcommand{\Th}{\hat{T}}
\renewcommand{\th}{\hat{t}}
\newcommand{\Wh}{\hat{W}}

\newcommand{\Hh}{\hat{H}}
\newcommand{\Sh}{\hat{S}}
\newcommand{\sh}{\hat{s}}

\newcommand{\rhoh}{\hat{\rho}}

\newcommand{\Gammah}{\hat{\Gamma}}

\usepackage[normalem]{ulem}
\usepackage{color}
\usepackage{xcolor}

\definecolor{Mygrey}{gray}{0.80}
\definecolor{lteal}{rgb}{0.10,0.60,0.70}
\definecolor{dkred}{rgb}{0.60,0.10,0.00}
\definecolor{Navy}{rgb}{0.00,0.00,0.60}
\definecolor{Magenta}{rgb}{0.94,0.20,0.90}
\definecolor{Green}{rgb}{0.24,0.71,0.29}
\newcommand{\SM}[1]{\textcolor{dkred}{SMSec~#1}}

\newcommand{\comment}[1]{}

\begin{document}

\title{Thermal Fundamental Gap Predictions in DFT via Optimally Tuned Hybrids}
%\title{Accurate finite-temperature fundamental gap predictions}

\author{Tim Gould}\email{t.gould@griffith.edu.au}
\affiliation{Qld Micro- and Nanotechnology Centre, %
  Griffith University, Nathan, Qld 4111, Australia}
\author{Leeor Kronik}
\affiliation{Department of Molecular Chemistry and Materials Science, Weizmann Institute of Science, Rehovoth 7610001, Israel}
\author{Gil Amoyal}
\affiliation{Department of Molecular Chemistry and Materials Science, Weizmann Institute of Science, Rehovoth 7610000, Israel}
\author{Stefano Pittalis}
\affiliation{Istituto Nanoscienze – CNR, S3, Via Campi 213A, I-41125 Modena, Italy}

\begin{abstract}
Predicting electronic fundamental gaps at finite temperature has remained conceptually and practically challenging.
We address this in three connected steps.
First, we extend generalized Kohn--Sham hybrid density functional theory to thermal ensembles, deriving a Mermin generalized Kohn--Sham framework from a thermal one-particle auxiliary system and an exact density-functional remainder.
Second, via an extension of Janak's theorem that holds rigorously in this framework, we recast Hirata's thermal-quasiparticle picture as a thermal orbital gap estimator and derive a closed low-temperature form, the error of which is controlled by the derivative discontinuity.
Third, because optimal tuning eliminates this error, the auxiliary orbital gap matches the interacting gap at low temperature, upgrading optimal tuning from a ground-state strategy to the governing principle -- mandatory, not optional -- for accurate finite-temperature gap predictions obtained from gaps of orbital eigenvalues within a hybrid functional framework.
We present applications that validate the theory and demonstrate its consequences.
\end{abstract}
%%%%%%%%%%%\pacs{31.15.ec,31.15.ep,03.65.Yz}
\maketitle

%AFQMC papers - He2019,Shen2020,Dornheim2026

Modern scientific and technological advances increasingly lead to regimes in which finite temperatures are exploited to control both nuclear and electronic behavior. Plasmas provide perhaps the most widespread example of high-temperature matter, with effective temperatures reaching tens of thousands of Kelvin and exhibiting an atypical separation between electronic and nuclear degrees of freedom. Plasmonic heating of nanoparticles offers another route to elevated (generally up to thousands of Kelvin) effective temperatures, with applications in cutting-edge quantum engineering \cite{Jauffred2019}. % Review on plasmonic heating

Modelling the behaviour of electrons in the warm dense matter regime (between condensed phase and plasma) is a rapidly advancing field~\cite{vorberger2025roadmapwarmdensematter}, building on pioneering work by Mermin~\cite{Mermin1965,Mermin1963-HF} as well as more recent advances in density functional theory (DFT)~\cite{Pittalis2011,Dufty2011,Dufty2015,Burke2016-FiniteT} and high-accuracy calculations~\cite{Blunt2015,Hummel2018,He2019,Shen2020,Dornheim2026} -- see also additional important work reviewed in Ref.\   \cite{vorberger2025roadmapwarmdensematter}.
Finite-temperature DFT provides a valuable balance between computational cost and accuracy, and has enabled a rapidly evolving understanding of matter under extreme conditions, whether through Kohn--Sham~\cite{KohnSham}-based approaches or orbital-free formulations~\cite{Karasiev2025}.
Recently, thermal density functional methods have been extended to address strongly quantum thermodynamic regimes as well \cite{Palamara2024,Palamara2025}.

Somewhat surprisingly, however, our ability to model the moderate-temperature regime (i.e. where the thermal physics is dominated by the lowest-lying charged and neutral excited states) remains less developed.
Despite recent foundational advances \cite{Hirata2024,Gu2024}, practical application of first-principles methods in this regime -- and the insight that can be gained from them -- remain limited.

With the goal of resolving this issue, this paper makes three connected advances.
First, we derive a generalized Kohn--Sham theory for finite temperatures via an extended Mermin formalism, providing the formal basis for hybrid and range-separated hybrid DFT in finite-temperature ensembles.
Second, building on Hirata's thermal-quasiparticle interpretation of orbital energies, we derive a closed low-temperature form of the thermal orbital finite-difference gap estimator and show that it is controlled by the derivative-discontinuity physics of the underlying hybrid.
Third, we upgrade optimal tuning~\cite{Livshits2007,Stein2009,Stein2010,Kronik2012,Wing2021} from a ground-state strategy to a governing principle for moderate-temperature gap predictions, without which high accuracy may be unattainable within a hybrid functional framework.

\emph{Mermin-Kohn-Sham DFT}:
Let us begin by defining the relevant many-electron problem.
The quantum mechanical grand canonical Gibbs free energy for electrons in an external potential $v$ and at temperature $\tau$ may be found via $G^{\tau}[v]=\min_{\Gammah}\tr[(\Hh[v]-\tau \Sh)\Gammah]$, where $\Hh[v]=\Th+\Wh+(\nh,v)$ is a Hamiltonian composed of a kinetic energy operator, $\Th$, an electron-electron interaction operator $\Wh$ and a potential operator $(\nh,v)$ that acts via the density; and $\Sh$ is an entropy operator~\footnote{Note, $\Tr[\Sh\Gammah]\propto -\Tr[\Gammah\log(\Gammah)]=-\sum_{\FE}w_{\FE}\log(w_{\FE})$.}.
The minimizing solution may be expressed as $\Gammah=\tfrac{1}{Z}e^{-(\Hh[v]-\mu \hat{N})/\tau}$ where  $\tau$ is the temperature, $\hat{N}$ is the electron number operator, $\mu$ is the chemical potential and $Z=\tr[e^{-(\Hh[v]-\mu \hat{N})/\tau}]\equiv \sum_{\FE}e^{-(E_{\FE}-\mu N_{\FE})/\tau}$ is the grand partition function.

% [not included] Blunt2014 10.1103/PhysRevB.89.245124 finite-T MC
% Hummel2018 10.1021/acs.jctc.8b00793 finite-T CC
% Blunt2015 10.1103/PhysRevLett.115.050603 finite-T MC
Despite recent advances in high-accuracy calculations~\cite{Blunt2015,Hummel2018,He2019,Shen2020,Dornheim2026} it remains very difficult to solve for $G$ or $\Gammah$ in many-electron systems.
DFT and approximations (DFAs) offer a way to bypass this difficulty at much more reasonable computational cost.
For thermal systems, Mermin-Kohn-Sham (MKS) theory~\cite{KohnSham,Mermin1965} is a popular DFT framework offering a good balance between cost and accuracy.
MKS theory can be defined as follows:
First, we re-express the free energy,
\begin{align}
    G^{\tau}[v]\equiv& \min_{n\to N}\big\{
    \Omega^{\tau}[n] + (n,v)
    \big\}\;.
    \label{eqn:FE}
\end{align}
in terms of a grand canonical density functional,
\begin{align}
    \Omega^{\tau}[n]:=&\min_{\Gammah\to n}\tr[(\Th + \Wh - \tau \Sh)\Gammah]\;.
    \label{eqn:Omega}
\end{align}%
Here, $\Gammah\to n$ indicates that $\tr[\nh\Gammah]=n$ and $n\to N$ indicates an average of $N$ electrons~\footnote{We write the equilibrium density simply as $n$ (and the 1RDM as $\rho$) rather than $n^{\tau}$ (and $\rho^{\tau}$).}.
Then we define a KS kentropy functional,
\begin{align}
\Omega_s^{\tau}[n]:=&\min_{\Gammah\to n}\tr[(\Th - \tau \Sh)\Gammah]\;.
\label{eqn:Omegas}
\end{align}
Finally, we incorporate any missing physics via the Hartree, exchange and correlation (Hxc) functional, $\Omega_{\Hxc}^{\tau}[n]:=\Omega^{\tau}[n]-\Omega_s^{\tau}[n]$, which can be approximated.
We may then find the free energy using $G^{\tau}[v]=\min_{n}\{\Omega_s^{\tau}[n] + \Omega_{\Hxc}^{\tau}[n]+(n,v)\}$.

Eq.~\eqref{eqn:Omegas} is especially useful because its minimizing argument may, without loss of generality, be constructed entirely from Slater determinant solutions of a \emph{non-interacting} Hamiltonian $\Th+(\nh,v_s)$.
In the absence of a magnetic field (assumed throughout) we may therefore treat each spin-orbital as an independent Fermion with an entropy $f_{i\sigma}\log(f_{i\sigma})+(1-f_{i\sigma})\log(1-f_{i\sigma})$ and kinetic energy $f_{i\sigma}t_i=f_{i\sigma}\ibraketop{\phi_i}{\th}{\phi_i}$ in the auxiliary problem 
\footnote{
The Slater determinants have energies $E_{s,\FE}=\ibraketop{\Phi_{\FE}}{\Th+(\nh,v_s)}{\Phi_{\FE}}=\sum_{i\sigma} \theta_{i\sigma}^{\FE}\epsilon_i=\sum_{i\sigma} \theta_{i\sigma}^{\FE}(t_i+v_{s,i})$ and electron numbers $N_{\FE}=\sum_{i\sigma} \theta_{i\sigma}^{\FE}$, where $\theta_{i\sigma}^{\FE}\in \{0,1\}$ indicates whether orbital $i$ with spin $\sigma$ is present in determinant $\iket{\FE_s}$, and $\epsilon_i$ is the spin-independent orbital energy.
It follows from the additivity of $N_{\FE}$ and $E_{s,\FE}$ that each system has a weight $w_{\FE}\propto \prod_{i\sigma\in\FE}\exp(-(\epsilon_i-\mu)/\tau)$ that is separable in $i$.
We may therefore replace sums over $\FE$ by products of sums over $\theta_{i\sigma}\in\{0,1\}$.}.
Crucially, we can define,
\begin{align}
    \Omega_s^{\tau}[n]:=&\min_{\rhoh_s\to n}\Omega_{1}^{\tau}[\rho_s]\;,
    &
    \Omega_{1}^{\tau}[\rho_s]:=&\tr[(\th - \tau\sh)\rhoh_s]
    \label{eqn:Omegas_rho}
\end{align}
where $\rhoh_s(\vr,\vrp)=\sum_i f_i\hat{\phi}_i^{\dag}(\vr)\hat{\phi_i}(\vrp)$ (for spin-summed $f_i=f_{i\up}+f_{i\down}$) is the one-body density matrix (1RDM) operator for the MKS system and $\rhoh\to n$ indicates that $\sum_i f_i |\phi_i|^2=n$.
Here, $\th$ and $\sh$ are one-body analogues of $\Th$ and $\Sh$ yielding $\tr[\th\rho]=\sum_i f_i t_i$ and $\tr[\sh\rhoh]= -2\sum_i  \{\tfrac{f_i}{2}\log(\tfrac{f_i}{2}) + (1-\tfrac{f_i}{2})\log(1-\tfrac{f_i}{2})\}$.

Minimizing the free energy subject to the density constraint yields KS orbitals and energies, along with occupation factors, obeying the mutual conditions
\begin{align}
&\big\{\th + v_s^{\tau}[n]\big\}\iket{\phi_i^{\tau}[n]}
=\epsilon_i^{\tau}[n]\iket{\phi_i^{\tau}[n]}\;,
\label{eqn:EKS}
\end{align}
and $n:=\sum_i f_i^{\tau}[n]|\phi_i^{\tau}[n]|^2$ where $v_s^{\tau}[n]:=v+\tfrac{\delta \Omega_{\Hxc}^{\tau}[n]}{\delta n}$ is the multiplicative KS potential and $f_i^{\tau}[n]:=2/[1 + e^{(\epsilon_i^{\tau}[n]-\mu^{\tau}[n])/\tau}]$ is the occupation factor for orbital $i$.
Finding the self-consistent field (SCF) solution of Eq.~\eqref{eqn:EKS} -- that is, iterating it until its input and output densities are the same -- minimizes the free energy when used in Eq.~\eqref{eqn:FE}.

Next, we are concerned with the structure of the Hxc functional and how best to approximate and use it for our aims.
Thermally averaging over Slater determinants yields,
\begin{align}
\Omega_{\Hx}^{\tau}[n]\equiv E_{\Hx}^{\HF}[n]=&E_{\Hrm}[n] + E_{\xrm}[\rho_s[n]]\;,
\label{eqn:Hx_rho}
\end{align}
where $E_{\Hrm}[n]:=\int n(\vr)n(\vrp)\tfrac{d\vr d\vrp}{2|\vr-\vrp|}$ and $E_{\xrm}[\rho]:=-\half\int |\rho(\vr,\vrp)|^2\tfrac{d\vr d\vrp}{2|\vr-\vrp|}$ are easy-to-evaluate ground state energy functionals.
The potential $v_{\Hrm}:=\tfrac{\delta E_{\Hrm}}{\delta n}$ is also easy to evaluate but $v_{\xrm}:=\tfrac{\delta E_{\xrm}}{\delta n}$ is difficult to evaluate~\cite{Grabo1997,Engel2003,Kummel2008}.

In ground states, the challenge of evaluating $v_{\xrm}$ motivated the development of generalized KS (GKS) theory~\cite{Seidl1996} which justifies~\cite{Goerling1997} hybrid approximations~\cite{Becke93} like PBE0~\cite{DFA:pbe1pbe} and B3LYP~\cite{Stephens1994}.
The key idea behind GKS theory is a subtle, yet important, adjustment to basic variables.
Rather than defining the 1RDM, $\rho_s[n]$, as a functional of the density, we instead express the density, $n[\rho_s](\vr)=\rho_s(\vr,\vr)$, as a functional of the 1RDM.
This simple change enables the inclusion of Fock exchange physics at low cost via a \emph{non-multiplicative} operator $\vh_{\xrm}$~\cite{Goerling1997}.
For ground state hybrid models it can be formally addressed by defining $Q^{\alpha}_{\GKS}[\rho]:=T_s[\rho] + E_{\Hrm}[n] + \alpha E_{\xrm}[\rho]$ as an auxiliary system invoking kinetic energy, Hartree and a fraction $\alpha$ of Fock exchange physics; and then defining $E_{\rxc}^{\alpha}[n]:=T_s[n]+E_{\Hxc}[n]-\min_{\rho\to n}Q^{\alpha}_{\GKS}[\rho]$ as a density functional capturing the missing exchange and correlations.
The SCF solution is then found via an effective potential $\vh_s=v+\alpha\vh_{\xrm}+v_{\rxc}^{\alpha}$ where $v_{\rxc}^{\alpha}=\tfrac{\delta E_{\rxc}^{\alpha}}{\delta n}$.

\emph{Generalization}:
We shall now proceed to show how Mermin GKS (MGKS) theory can be rigorously motivated.
The main challenge to overcome is that hybrid solutions are not strictly non-interacting, which has a major impact on the definition of the entropy and free energy;~\cite{Gu2024,Hirata2024} crucially they lack the convenient separable form of pure MKS theory that leads to Eq.~\eqref{eqn:Omegas_rho}.
One solution to this problem is to adapt a free energy functional~\cite{Gu2024} and then use ensemble GKS (EGKS) theory to obtain the variational minima~\cite{Gould2021-EGKS}.
However, this leads to severe practical complications.

Here, we adopt a more direct solution that avoids such complications: we use the MKS 1RDM functionals [Eq.~\eqref{eqn:Omegas_rho} and \eqref{eqn:Hx_rho}] as building blocks for the Mermin GKS (MGKS) system 
\begin{align}
Q^{\alpha,\tau}_{\MGKS}[\rho]:=&\Omega_1^{\tau}[\rho] + E_{\Hrm}[n] + \alpha E_{\xrm}[\rho]\;,
\label{eqn:G_rho}
\end{align}
and restore the remainder physics via
\begin{align}
\Omega_{\rxc}^{\alpha,\tau}[n]:=&\Omega^{\tau}[n]
- \min_{\rho\to n} Q^{\alpha,\tau}_{\MGKS}[\rho]\;.
\label{eqn:Omega_rxc_n}
\end{align}
Here, $Q_{\MGKS}^{\alpha,\tau}$ defines the auxiliary MGKS reference system: a thermal one-particle ensemble with the MKS entropy and a non-multiplicative exchange term.
The remainder functional $\Omega_{\rxc}^{\alpha,\tau}$ is then defined by subtraction, so that the total constrained search recovers the exact Mermin functional.
It follows that
\begin{align}
G^{\tau}[v]
=&\min_{n\to N} \big\{ \min_{\rho\to n} Q^{\alpha,\tau}_{\MGKS}[\rho] + \Omega_{\rxc}^{\alpha,\tau}[n] + (n,v) \big\}\;,
\label{eqn:Gibbs_rho}
\end{align}
is independent of $\alpha$, by construction.

Our next step is to derive the minimizing solution of Eq.~\eqref{eqn:Gibbs_rho}.
Using $\min_{n\to N}\min_{\rho\to n}=\min_{\rho\to N}$ and expanding $Q_{\MGKS}^{\alpha,\tau}[\rho]$ and $\Omega_{\rxc}^{\alpha,\tau}[n]$ yields
$G^{\tau}[v]
=\min_{\rho\to N}\big\{ T_1[\rho] + \alpha E_{\xrm}^{\HF}[\rho] - \tau S_1[\rho] 
+ E_{\Hrm}[n] + \Omega_{\rxc}^{\alpha,\tau}[n] + (n,v) \big\}$,
where we implicitly use $n(\vr)=\rho(\vr,\vr)$ in later terms.
Here, $T_1[\rho]:=\tr[\th\rhoh]$ and $S_1[\rho]:=\tr[\sh\rhoh]$ are components of $\Omega_1^{\tau}[\rho]$.
We next expand $\rho(\vr,\vrp)=\sum_i f_i \varphi_i(\vr)\varphi_i^*(\vrp)$; and introduce the constraints $\ibraket{\varphi_i}{\varphi_j}=\delta_{ij}$ and $\sum f_i=N$ with Lagrange multipliers $f_i\varepsilon_{ij}$ and $\mu$, respectively.
The free energy is therefore minimized when,
\begin{align}
0=&f_i[ \th + v + \vh_{\Hxc}^{\alpha,\tau} ] \ket{\varphi_i} - \sum_j f_i \varepsilon_{ij} \iket{\varphi_j}\;,
~~\forall i,
\label{eqn:L_phi}
\\
0=&\ibraketop{\varphi_i}{\th + v + \vh_{\Hxc}^{\alpha,\tau}}{\varphi_i} -\tau \tfrac{\partial S_1}{\partial f_i} - \mu
\label{eqn:L_f}
\end{align}
for variations w.r.t. $\varphi_i$ and $f_i$ respectively, where we introduced $\vh_{\Hxc}^{\alpha,\tau}=\tfrac{\delta [E_{\Hrm} + \alpha E_{\xrm} + \Omega_{\rxc}^{\alpha,\tau}]}{\delta\rhoh}\equiv v_{\Hrm} + \alpha \vh_{\xrm}+v_{\rxc}^{\alpha,\tau}$.

Finally, recognizing that $\vh_{\Hxc}^{\alpha,\tau}$ is Hermitian lets us set $\varepsilon_{ij}=\delta_{ij}\epsilon_i^{\alpha}$ so that Eq.~\eqref{eqn:L_phi} becomes,
\begin{align}
\big\{\th + \vh_s^{\alpha,\tau}[\rho]\big\}\iket{\phi_i^{\alpha,\tau}[\rho]}
=\epsilon_i^{\alpha,\tau}[\rho]\iket{\phi_i^{\alpha,\tau}[\rho]}\;,
\label{eqn:EGKS}
\end{align}
i.e. similar to Eq.~\eqref{eqn:EKS} but with density functionals replaced by 1RDM functionals.
Using $\ibraketop{\phi_i}{\th+v+\vh_{\Hxc}^{\alpha}}{\phi_i}=\epsilon_i$ and $\tfrac{\partial S_1}{\partial f_i}=-\log\tfrac{f_i}{2-f_i}$ in Eq.~\eqref{eqn:L_f} yields,
\begin{align}
%\epsilon_i^{\alpha}+\tau \log\tfrac{f_i^{\alpha}}{2-f_i^{\alpha}}-\mu=0
%~~\Rightarrow~~
f_i^{\alpha,\tau}[\rho]=&\frac{2}{1 + e^{(\epsilon_i^{\alpha,\tau}[\rho]-\mu[\rho])/\tau}}\;,
\label{eqn:fi}
\end{align}
again similar to the ensemble Kohn-Sham case.
Importantly, the MGKS orbital energies obey \footnote{Applying $\tfrac{\partial F[\rho]}{\partial f_i}=\ibraketop{\phi_i}{\tfrac{\delta F}{\delta\rho}}{\phi_i}$ to the auxiliary enthalpy functional $E^{\alpha,\tau}[\rho]=T_1[\rho]+E_{\Hrm}[n]+\alpha E_{\xrm}[\rho] + \Omega_{\rxc}^{\alpha,\tau}[n] + (n,v)$ gives $\tfrac{\partial E^{\alpha,\tau}}{\partial f_i}=\ibraketop{\phi_i^{\alpha,\tau}}{\th + v + \vh_{\Hxc}^{\alpha,\tau}}{\phi_i^{\alpha,\tau}}\equiv \epsilon_i^{\alpha,\tau}$.}
a thermal extension, $\tfrac{\partial E^{\alpha,\tau}}{\partial f_i}=\epsilon_i^{\alpha,\tau}$, of Janak's theorem~\cite{Janak1978}.

Crucially, each $\alpha$ yields its own exact MGKS hybrid [Eqs.~\eqref{eqn:G_rho} and \eqref{eqn:Omega_rxc_n}] and its own SCF solution [Eqs~\eqref{eqn:EGKS} and \eqref{eqn:fi}].
Auxiliary system properties (e.g., orbitals, orbital energies, occupation factors, KS enthalpy and entropy) are defined by the SCF solution, and therefore vary with $\alpha$.
That is, while the `true' interacting free energy [Eq.~\eqref{eqn:Gibbs_rho}] is independent of $\alpha$, we have multiple treatments for obtaining the exact solution.
MGKS theory may also be extended to range-separated hybrids (RSH), in which the separation depends on $\alpha$ plus a range-separation parameter $\omega$ and long-range constant $\alpha_{\text{lr}}$ (that we leave out of superscripts for notational brevity), as done in Supplementary Material Section~I (\SM{I}) 
\footnote{See Supplemental Material at [URL will be inserted by publisher] for further analysis and technical details of calculations; including references~\cite{Perdew1981,Gould2022-HL,Gould2025-GX24,Gould2026-Perspective}.}.

\emph{Behaviour of the MGKS system}:
For ground-state DFT, key auxiliary properties of KS and GKS approaches are typically very similar
%10.1103/PhysRevB.68.035103
~\cite{Kummel2003,Garrick2020,Garrick2022}.
In a two electron ground state these similarities become {\em identities} because using a non-multiplicative GKS x potential is equivalent to using a multiplicative KS potential $v_{\xrm}=-\half v_{\Hrm}$.
It follows from similarity that the remainder xc energy may be approximated as $E_{\rxc}^{\alpha,\text{GKS}}\approx (1-\alpha)E_{\xrm}^{\HF} + E_{\crm}^{\KS}$ with minimal loss of accuracy; and it is this relationship that underpins the practical success of hybrid approximations, where the exchange and correlation terms in $E_{\rxc}^{\DFA}\approx (1-\alpha)E_{\xrm}^{\DFA} + E_{\crm}^{\DFA}$ are replaced by DFAs.

The `virtual' orbitals (i.e., orbitals that are unoccupied at 0~K) vary significantly with $\alpha$ (and $\omega$, $\alpha_{\text{lr}}$ for RSH), even for two-electron systems, because non-multiplicative and multiplicative exchange potentials have different asymptotic impacts on virtual orbitals.
In ground states, the main impact of interest is on the relationship between frontier orbital energy differences and exact properties.
The ionisation potential theorem \cite{Perdew1982,Almbladh1985} dictates that the highest occupied orbital energy obeys $\epsilon_h^{\alpha}=E^{N}-E^{N-1}$ regardless of the value of $\alpha$.
The first virtual orbital energy may be written as $\epsilon_l^{\alpha}=E^{N+1}-E^{N}-\Delta_{\xc}^{\alpha}$ where the additional derivative-discontinuity~\cite{Perdew1982} term, $\Delta_{\xc}^{\alpha}$, is a function of $\alpha$ (and $\omega$, $\alpha_{\text{lr}}$) and a functional of $n$.
Here, $\Delta_{\xc}^{\alpha}$ denotes the standard derivative-discontinuity contribution to the fundamental gap.
It follows that $\Delta := E^{N+1}+E^{N-1}-2E^N = \Delta_s^{\alpha}+\Delta_{\xc}^{\alpha}$ where 
$\Delta_s^{\alpha}:=\epsilon_l^{\alpha}-\epsilon_h^{\alpha}$.
This relation is subject to some order of limits issues; throughout we assume the order in which application of ground state theory (no $\tau$ superscripts) is equivalent to the 0~K thermal limit.

In the ground state the above  orbital-gap relations are tied to Janak's theorem~\cite{Janak1978}.
Hirata's thermal-quasiparticle picture invokes Janak-like assumptions -- given rigorous footing below Eq.~\eqref{eqn:fi} -- to motivate orbital energy-based thermal gap descriptions~\cite{Hirata2024}.
Following Hirata, we construct an orbital energy $E_s^{\alpha,\tau}(M)=\sum_i f_{i;M}^{\alpha,\tau}\epsilon_i^{\alpha,\tau}$ for an $M$-electron system by holding the orbital energies $\epsilon_i^{\alpha,\tau}$ fixed at those of the neutral $N$-electron system and varying only the Fermi--Dirac occupations [i.e. $\mu\to M$, not $N$, in Eq.~\eqref{eqn:fi}].
Its second finite difference in $M$ yields the thermal orbital gap estimator
\begin{align}
\Delta_s^{\alpha,\tau}=&\sum_i [f_{i;N+1}^{\alpha,\tau}+f_{i;N-1}^{\alpha,\tau}-2f_{i;N}^{\alpha,\tau}]
\epsilon_i^{\alpha,\tau}\;,
\label{eqn:HGtau}
\end{align}
which extends the usual 0~K frontier-orbital picture to finite temperatures; the full derivation and extensions are given in \SM{II}.

Eq.~\eqref{eqn:HGtau} can be read as the finite-temperature analogue of using frontier orbital energies to estimate an addition/removal gap.
The key observation to proceed is to notice that Eq.~\eqref{eqn:HGtau} obeys $\Delta_s^{\alpha,\tau\to 0}=\epsilon_l^{\alpha}-\epsilon_h^{\alpha}=\Delta - \Delta_{\xc}^{\alpha}$, yet embeds richer physics at higher temperatures.
To obtain a simple low-temperature form, we assume an even, closed-shell $N$ and retain only the lowest states in the $N$, $N-1$, and $N+1$ charge sectors.
This restricts the thermal response to the frontier occupations and is only appropriate when $\tau\ll \Delta$~\footnote{Note, this is more restrictive than the general moderate temperature case.
More detailed assumptions are discussed in \SM{II}.}.
To leading low-temperature order we obtain $f_l^{\alpha,\tau}\approx 2/[1+e^{(\epsilon_l^{\alpha}-\epsilon_h^{\alpha})/(2\tau)}]=2-f_h^{\alpha,\tau}$ so that,
\begin{align}
    \Delta_s^{\tau,\alpha}\approx \; &
    \Phi^{\tau}\big(\epsilon_l^{\alpha}-\epsilon_h^{\alpha}\big)
    =
    \Phi^{\tau}\big(\Delta-\Delta_{\xc}^{\alpha}\big)
    \;,
    \label{eqn:HGtau_lowT}
\end{align}
where  
$\Phi^{\tau}(\omega):=\omega/[1 + 4e^{-\omega/(2\tau)}]$.

Next, the crucial question: How does Eq.~\eqref{eqn:HGtau_lowT} compare to the {\em interacting} finite-temperature gap, $\Delta^{\tau}$?
Here, $\Delta^{\tau}:=E^{\tau,N+1}+E^{\tau,N-1}-2E^{\tau,N}$ is defined as the finite difference of thermally averaged interacting $M$-electron energies $E^{\tau,M}=\sum_{\FE}E_{\FE}e^{-(E_{\FE}-\mu^M N_{\FE})/\tau}/Z^{\tau,\mu^M}$, where $Z^{\tau,\mu} = \sum_{\FE}e^{-(E_{\FE}-\mu N_{\FE})/\tau}$ is the interacting grand partition function and $\mu^M$ ensures that the average particle number obeys $M=\sum_{\FE}N_{\FE}e^{-(E_{\FE}-\mu^M N_{\FE})/\tau}/Z^{\tau,\mu^M}$.
Using the same low-temperature approximation as above yields $\Delta^{\tau}\approx \Phi^{\tau}(\Delta)$.

Comparing the two low-temperature forms, $\Delta_s^{\alpha,\tau}\approx\Phi^{\tau}(\Delta-\Delta_{\xc}^{\alpha})$ and $\Delta^{\tau}\approx\Phi^{\tau}(\Delta)$, shows that the orbital estimator reproduces the interacting gap precisely when the derivative discontinuity $\Delta_{\xc}^{\alpha}$ vanishes.
The final step is therefore to eliminate $\Delta_{\xc}^{\alpha}$ using a ground-state DFT strategy -- optimally-tuned (OT) hybrids~\cite{Livshits2007,Stein2009,Stein2010,Kronik2012} -- that is founded on the result that there are generally combinations of $\alpha_o$, $\omega_o$ and $\alpha_{\text{lr},o}$ for which $\Delta_{\xc}^{\alpha_o}=0$ so that $\Delta_s^{\alpha_o}=\Delta$.
Substituting $\Delta_{\xc}^{\alpha_o}=0$ into Eq.~\eqref{eqn:HGtau_lowT} then renders it identical to its interacting counterpart $\Phi^{\tau}(\Delta)$, and ensures that the traditional 0~K gap OT equality extends to the low-temperature regime.
Further details and analysis of theory are provided in \SM{II}.

Importantly, the success of the MGKS construction and Eq.~\eqref{eqn:HGtau_lowT} offer an effective strategy for predicting temperature dependent fundamental gaps when reference calculations are intractable.
OT-RSH functional \emph{approximations} are constructed~\cite{Kronik2012} to give $\Delta_{\xc}^{\text{OT-RSH}}\approx 0$, and generally enforce $\Delta\approx \epsilon_l-\epsilon_h$ quite accurately~\cite{Kronik2012,Wing2021}.
Eq.~\eqref{eqn:HGtau_lowT} ensures that this 0~K success can translate to low temperatures.
Thus, we expect that an OT-RSH will outperform other approximations across a range of temperatures and thereby provide a pragmatic reference for studying complex systems and/or testing popular DFAs for realistic systems.

\emph{Applications}:
The principle above is established [via Eq.~\eqref{eqn:HGtau_lowT}] on general theoretical grounds.
The numerical examples that follow validate it against exact results (He) and then demonstrate consequences on realistic systems.
In the applications below, the OT-RSH parameters are fixed by the standard 0~K tuning condition and then used unchanged in Eq.~\eqref{eqn:HGtau}.
Note, as our goal is to understand the temperature-effects of different MGKS treatments we avoid using thermally-adapted DFAs~\cite{Karasiev2014,Groth2017,Karasiev2018,Hilleke2025} whose $\tau$-dependence would complicate the analysis; and preserve a fixed nuclear geometry for the same reason.
Calculations are carried out using Psi4~\cite{Smith2018,Smith2020} (He, Li clusters) and VASP~\cite{Kresse1993,Kresse1994,Kresse1996} (solid Ge).

%%%%%%%%%%%%%%%%%%%%%%%%%%%%%%%%%%%%%%%%%%%%%%%%%
\begin{figure}[t!h]
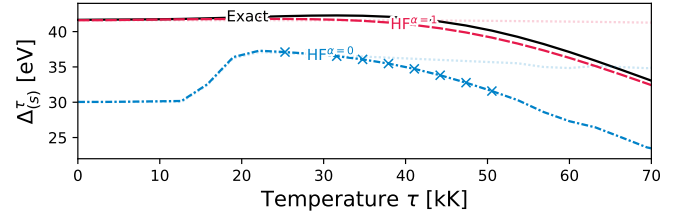

    \includegraphics[width=\linewidth]{{{Fig_He_Simple}}}
    \caption{Finite-temperature gap in He computed from exact interacting solutions (black line) and from Hartree-Fock MGKS theory using $\alpha=0$ (blue dash-dot) and $\alpha=\alpha_o=1$ (red dashes).
    Crosses indicate interpolated values.}
    \label{fig:He}
\end{figure}
%%%%%%%%%%%%%%%%%%%%%%%%%%%%%%%%%%%%%%%%%%%%%%%%%

We first validate the theory on a simple yet illuminating case:
atomic He (in a harmonic well to ensure a good thermodynamic limit), where exact reference results are provided by wave function theory.
Figure~\ref{fig:He} applies the Hartree-Fock approximation (i.e. $E_{\crm}=0$) to MGKS theory with $\alpha=0$ and $\alpha=1$; and compares with exact results.
The former case yields a significant derivative-discontinuity term $\Delta_{\xc}^0\approx 11.6$~eV and fails, quantitatively and qualitatively, to reproduce exact behaviors at any temperature.
By contrast, $\alpha=1=\alpha_o$ yields $\Delta_{\xc}^1\approx 0$ so is an OT-hybrid; and we see it also yields excellent agreement across all reported temperatures.
Figure~\ref{fig:He} thus provides a direct validation of the theory.
Detils and data are in \SM{III}.

%%%%%%%%%%%%%%%%%%%%%%%%%%%%%%%%%%%%%%%%%%%%%%%%%
\begin{figure}[b!h]
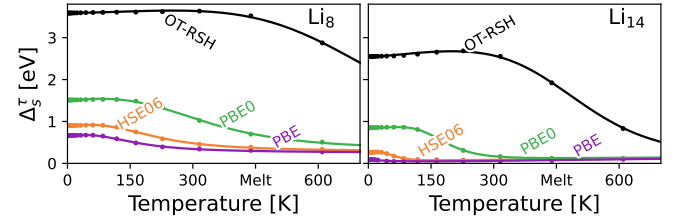

\includegraphics[width=\linewidth]{{{Fig_Li_SmearedGap}}}
\caption{Eq.~\eqref{eqn:HGtau} applied to small Li clusters using PBE (purple), PBE0 (green), HSE06 (orange) and OT-RSH (black).
\label{fig:LiGap}
}
\end{figure}
%%%%%%%%%%%%%%%%%%%%%%%%%%%%%%%%%%%%%%%%%%%%%%%%%

Next, to demonstrate the consequences of the theory and the limitations of current DFAs, we investigate small Li clusters with 8 and 14 atoms at fixed geometry.
We apply Eq.~\eqref{eqn:HGtau} in the PBE, PBE0 hybrid, and HSE06 range-separated hybrid approximations. We use an OT-RSH as reference -- as justified by Eq.~\eqref{eqn:HGtau_lowT} and supported (see \SM{IV.C}) by a strong performance of OT-RSH on Li$_2$.
Crucially, Figure~\ref{fig:LiGap} reveals that different DFAs yield dramatically different quantitative and qualitative trends at low temperatures.
The reference OT-RSH calculation predicts a large gap that grows at very low temperatures, before decreasing -- a feature also seen in HSE06 but not in PBE or PBE0.
Details and data are in \SM{IV}.

%%%%%%%%%%%%%%%%%%%%%%%%%%%%%%%%%%%%%%%%%%%%%%%%%
\begin{figure}[t!h]
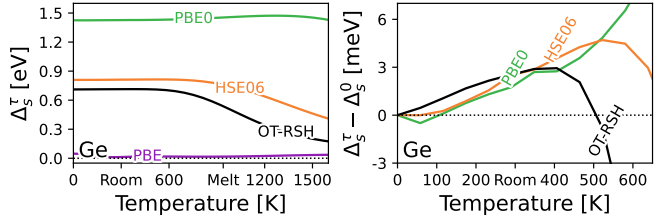

\includegraphics[width=\linewidth]{{{Fig_Ge}}}
\caption{Left: Eq.~\eqref{eqn:HGtau} applied to solid Ge using  PBE (purple), PBE0 (green), HSE06 (orange) and OT-RSH (black).
Right: $\Delta\Delta_s^{\tau}=\Delta_s^{\tau}-\Delta_s^{0}$ at low temperatures.
\label{fig:GeGap}
}
\end{figure}
%%%%%%%%%%%%%%%%%%%%%%%%%%%%%%%%%%%%%%%%%%%%%%%%%

Finally, we investigate a solid-state system.
Figure~\ref{fig:GeGap} shows that solid crystalline Ge exhibits similar behaviour to the Li clusters, with both the 0~K gap and its low-temperature changes sensitive to the chosen DFA.
Ge is a stringent test because its FG (0.7~eV) is very small; and PBE predicts zero gap.
In this case, all hybrid approximations predict a slight initial increase in the gap; but the transition to a decrease happens at much lower temperatures in OT-RSH compared to PBE0 (outside the figure) and HSE06.
Note, PBE strictly gives no gap so we exclude it from the $\Delta\Delta_s$ plot.
Details and data, plus extension of theory to periodic systems, are in \SM{V}.

\emph{Conclusions}:
To conclude, this work makes three connected advances.
First, we formulate a Mermin generalized Kohn--Sham framework by defining a thermal 1RDM auxiliary system [Eq.~\eqref{eqn:G_rho}] and an exact density-functional remainder [Eq.~\eqref{eqn:Omega_rxc_n}].
This provides the formal basis for applying hybrid and range-separated hybrid DFT to finite-temperature ensembles.
Second, building on Hirata's thermal-quasiparticle interpretation, we replace the zero-temperature HOMO/LUMO proxy by the thermal orbital finite difference of Eq.~\eqref{eqn:HGtau}, which preserves the 0~K limit while incorporating thermal occupations.
Third, we show that the low-temperature behaviour of this estimator is controlled by the same derivative-discontinuity physics that motivates optimal tuning at 0~K [Eq.~\eqref{eqn:HGtau_lowT}].
Thus, optimal tuning is not merely a ground-state correction: within the present framework, it provides a direct route to reliable low- and moderate-temperature gap predictions.

Concerning the applications:
The case of He (Figure~\ref{fig:He}) validates the theory behind Eq.~\eqref{eqn:HGtau} because optimal $\alpha=1$ yields excellent agreement with reference calculations.
Eq.~\eqref{eqn:HGtau} reveals strong DFA sensitivity in Li clusters and solid Ge (Figures~\ref{fig:LiGap} and \ref{fig:GeGap}), showing that finite-temperature gap trends from HOMO/LUMO gaps need to be interpreted with caution.

At higher temperatures, the present 0~K-based DFAs should ultimately be combined~\cite{Mihaylov2020,Ellaboudy2025} with thermally adapted exchange-correlation free energies, so that optimal tuning controls the low-temperature gap while explicit entropic terms control the high-temperature limit.
By suppressing the derivative-discontinuity error, OT-RSH provides a pragmatic reference for the moderate-temperature regime and establishes optimal tuning, on theoretical grounds, as the route for extending successful 0~K gap strategies to thermal electronic excitations.

\vspace{5mm}

\acknowledgments
TG and LK were supported by an Australian Research Council (ARC) Discovery Project (DP200100033).
TG was supported by an ARC Future Fellowship (FT210100663). LK was supported by the Aryeh and Mintzi Katzman Professorial Chair and the Helen and Martin Kimmel Award for Innovative Investigation.
Computing resources were provided by the Australian National Computing Merit Application Scheme (NCMAS sp13).
TG would like to thank Nathan Garland for useful discussions about plasmas.

\bibliography{EDFT}

\end{document}